\documentclass[conference]{IEEEtran}
\IEEEoverridecommandlockouts
\usepackage{cite}
\usepackage{amsmath,amssymb,amsfonts,amsthm}
\usepackage{algorithmic}
\usepackage{graphicx}
\usepackage{textcomp}
\usepackage{xcolor}
\usepackage{bm}
\usepackage{subcaption}
\usepackage{mathtools} % Bonus

\theoremstyle{definition}
\newtheorem{theorem}{Theorem}

\graphicspath{{figure/}}
\def\BibTeX{{\rm B\kern-.05em{\sc i\kern-.025em b}\kern-.08em
    T\kern-.1667em\lower.7ex\hbox{E}\kern-.125emX}}
\begin{document}

%\title{Physical Layer Security-Assisted Computation Offloading in Blockchain-Empowered Internet of Things -- A Gas-Oriented Approach

\title{Endogenous Security of Computation Offloading in Blockchain-Empowered Internet of Things

}

\author{\IEEEauthorblockN{Yiliang Liu}
\IEEEauthorblockA{\textit{School of Cyber Science and Engineering} \\
\textit{Xi'an Jiaotong University}\\
Xi'an, China \\
{\tt liuyiliang@xjtu.edu.cn}}
\and
\IEEEauthorblockN{Zhou Su}
\IEEEauthorblockA{\textit{School of Cyber Science and Engineering} \\
\textit{Xi'an Jiaotong University}\\
Xi'an, China \\
{\tt zhousu@ieee.org}}
\and
\IEEEauthorblockN{Bobo Yu}
\IEEEauthorblockA{\textit{Department of Engineering} \\
\textit{Nanjing Kean Electronics Co., Ltd}\\
Nanjing, China \\
{\tt yubobo\_1990@126.com}}

}

\maketitle

\begin{abstract}
This paper investigates an endogenous security architecture for computation offloading in the Internet of Things (IoT), where the blockchain technology enables  the traceability of malicious behaviors, and the task data uploading link from sensors to small base station (SBS) is protected by intelligent reflecting surface (IRS)-assisted physical layer security (PLS). After receiving task data, the SBS allocates computational resources to help sensors perform the task. The existing computation offloading schemes usually focus on network performance improvement, such as energy consumption minimization, and neglect the Gas fee paid by sensors, resulting in the discontent of high Gas payers. Here, we design a Gas-oriented computation offloading scheme that guarantees the degree of satisfaction of sensors, while aiming to reduce energy consumption. Also, we deduce the ergodic secrecy rate of IRS-assisted PLS transmission that can represent the global secrecy performance to allocate computational resources. The simulations show that the proposed scheme ensures that the node paying higher Gas gets stronger computational resources, and just raises $4\%$ energy consumption in comparison with energy consumption minimization schemes.
\end{abstract}


\section{Introduction}
With the rapid popularization of the Internet of Things (IoT) technologies in industrial manufacture, business, etc., the endogenous security issues of heterogeneous devices and infrastructures have raised many concerns. The endogenous security issues of IoT networks focus on passive eavesdropping attacks and unknown malicious behaviors of nodes \cite{Jin2021}. Nowadays, physical layer security and blockchain technologies are regarded as promising endogenous security technologies to achieve the traceability of malicious behaviors and confidentiality of IoT networks, respectively \cite{Wang2021,He2021,Qiu2021,Wang2021a,Yin2021}.

One of the typical applications in IoT is the computation offloading of sensor nodes, where the endogenous security architecture is established to secure the offloading process \cite{Wang2021a,Yin2021}. Especially, smart contracts record the offloading process participated by sensor nodes and mobile edge computing (MEC) servers in the blockchain to ensure that transactions cannot be denied and malicious computing result providers will also be traced. IRS-assisted PLS can control the direction of the electromagnetic wave to improve the secrecy capacity between sensors and MEC servers. However, the existing computational resource allocation schemes are not suitable in the architecture. Firstly, existing schemes usually focus on the optimization of system performance, such as latency and energy consumption minimization \cite{Wang2016,Feng2020,Liu2021,Du2018}. The Gas\footnote{Gas refers to the cost necessary to perform transactions on the Ethereum blockchain} factor just affects the block generation speed and is not considered in computational resource allocation, leading to the dissatisfaction of sensors because better resources are not allocated to those sensors even if they pay high Gas. In addition, existing investigation on IRS-assisted PLS neglects the expression of ergodic secrecy rates \cite{Sai2021}, resulting in an inappropriate resource allocation in time-varying wireless channels.
 
To address these problems mentioned above, we present Gas-oriented computational resource allocation to guarantee that the node paying higher Gas has more opportunities to get a stronger computational resource. Also, the expression of ergodic secrecy rate is deduced for the allocation process. The main contributions of this work are summarized as follows.
\begin{itemize}
\item We formulate an endogenous security architecture for computation offloading in  IoT, where IRS-assisted PLS and blockchain achieve the link security and traceability of malicious behaviors, respectively.

\item We deduce the expression of ergodic secrecy rate of IRS-assisted PLS channels, which provides a global metric of secrecy performance for the computational resource allocation process.

\item We design a Gas-based computational resource allocation algorithm where sensors are divided into different groups based on Gas payments, and the group with higher Gas is prioritized with better computational resources.
\end{itemize}

The remainder of the paper is organized as follows. Section \ref{model} describes the system model. The computation offloading scheme is proposed in Section \ref{proposed1}. We show simulation results in Section \ref{simulations}, and conclude this paper in Section \ref{conclusions}.

%\textsl{Notations:} Bold uppercase letters, such as $\mathbf{A}$, denote matrices, and bold lowercase letters, such as $\mathbf{a}$, denote column vectors. $\mathbf{A}^{\rm{H}}$ represents the conjugate transpose of $\mathbf{A}$. $\mathbf{I}_a$ is an identity matrix with its rank $a$. $\mathcal{CN}(\mu,\sigma^2)$ is a complex normal (Gaussian) distribution with mean $\mu$ and variance $\sigma^2$. $|\mathbf{x}|$ is the Euclidean norm of $\mathbf{x}$. $\text{diag}(\mathbf{X})$ is the diagonal matrix of $\mathbf{X}$. $\mathbb{E}(\cdot)$ is the expectation operation. $\text{arg}(x)$ is the angle of complex variable $x$. $\lceil x \rceil$ is the ceil of $x$.

%where the sensor $U_i$ sends computation tasks to the SBS via IRS and instantaneously broadcasts contracts of publishing tasks denoted by $\#1$. After receiving tasks, the SBS allocates computational resources of MEC server $M_k$ to complete the tasks, then broadcasts contracts of recording results denoted by $\#2$. Both task publishing and results uploading should be recorded in blocks via the $\#1$ block generation and $\#2$ block generation, respectively. The uplink links from sensors to the SBS is protected by PLS technologies. 

\section{System Model and Problem Formulation}\label{model}
\begin{figure}[!htp]
\centering
\includegraphics[width=1\linewidth]{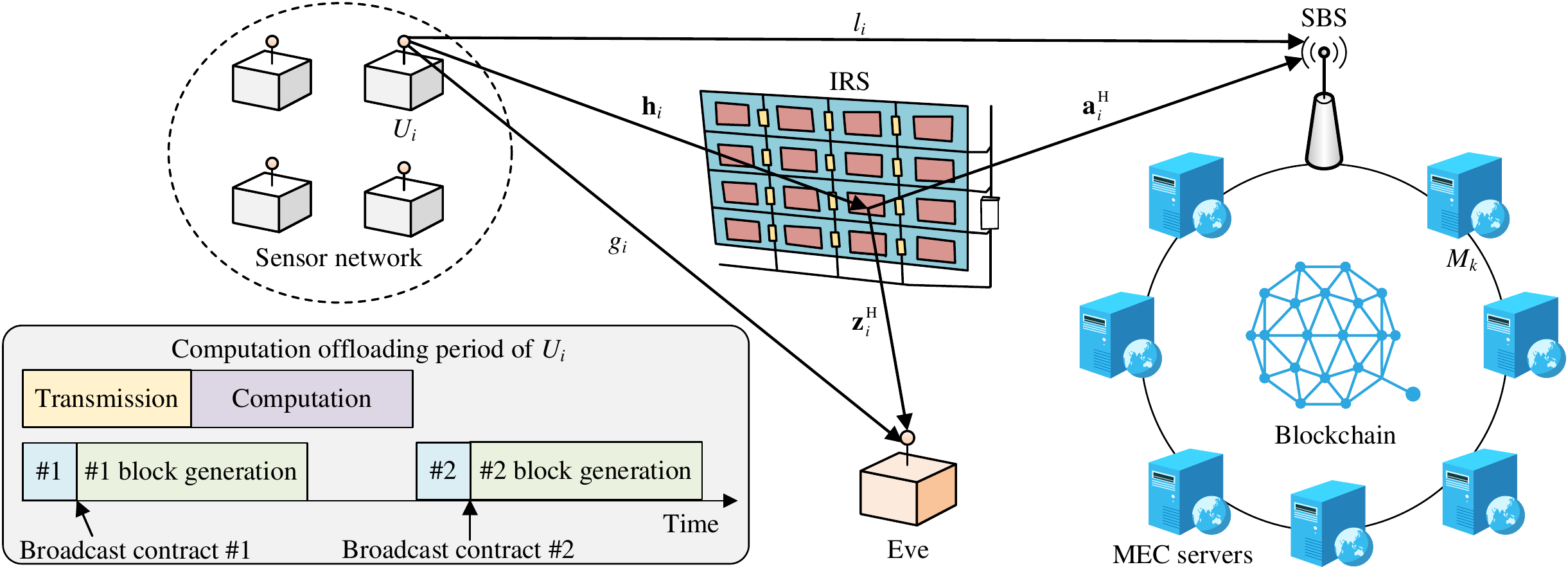}
\caption{Joint physical layer security and blockchain-assisted computation offloading in IoT networks}\label{model_figure}
\end{figure}

This article considers a blockchain-empowered IoT network, as shown in Fig. \ref{model_figure}, which includes $N_I$ sensor nodes, denoted by $\{U_1,U_2,...,U_{N_I}\}$. Each sensor is equipped with one antenna. A set of $N_K$ MEC servers, denoted by $\{M_1,M_2,...,M_{N_K}\}$, are associated with a single-antenna SBS via wired links, and $N_K \geq N_I$. All sensors and MEC servers are registered in a public Ethereum network and follow the rules of the Ethereum network. Due to the transmission power constraints and the large-scale fading effect of sensor communications, an IRS device is deployed to enhance the quality of the uplink channel from sensors to the SBS.

As the computational resources of sensors are scarce, these $N_I$ sensors offload their computational tasks, denoted by $\{Z_1,Z_2,...,Z_{N_I}\}$, to MEC servers via the wireless channels between sensors and the SBS. After receiving tasks, the SBS allocates virtual machines provided by MEC servers to execute these tasks. To record these tasks, sensors store the indices of publishing tasks on the Ethereum via a contract function, i.e., $\mathsf{task\_publish\_contract}(\cdot)$, and the SBS stores the indices of results on the Ethereum via the contract function $\mathsf{result\_record\_contract}(\cdot)$. Once the blocks of these contracts are deployed on the Ethereum, no one can repudiate the debts of transactions, and the results given by MEC servers are recorded to avoid the attacks of malicious MEC servers. The computation offloading process in the blockchain-empowered IoT network is described as follows. 
\begin{enumerate}
\item Transmitting tasks to the SBS: $U_i$ sends the task data to the SBS via IRS-assisted PLS schemes to resist eavesdropping attacks. To avoid the interference, sensors use the time-division multiple access (TDMA) technology during transmission processes.
\item Creating contracts of publishing tasks: At the same time of task transmission, by following the contract format of Ethereum, the sensor $U_i$ launches a contract, i.e., $\#1=\mathsf{task\_publish\_contract}(Z_i)$, which includes the hash message corresponding to the offloaded task $Z_i$ and the signature of $U_i$. Also, the contract has the address information of the transaction parties, Gas for this transaction defined by $V_i$, Gas price, etc. Then, the contract $\#1$ is broadcast in Ethereum.
\item Computing tasks in MEC server: After receiving tasks, the SBS allocates computational resources of MEC servers to perform these tasks. It is assumed that a task uses one MEC server to compute results and an MEC server can be allocated to only one task. At last, the results are uploaded to the SBS for subsequent services.
\item Generating blocks of contracts of publishing tasks: All nodes in Ethereum synchronize the transaction of $\mathsf{task\_publish\_contract}(Z_i)$, and check its format and signature. If passing, nodes compete with each other to win the right of charging the account of the transaction, then the block is generated in Ethereum by winners. All members registered in Ethereum desiring to get payments can take part in the competition of account-charging rights.
\item Creating contracts of recording results: Without loss of generality, we assume that the task of $U_i$, i.e, $Z_i$, is offloaded to the MEC server $M_k$. After completing the task, $M_k$ launches a contract, i.e., $\mathsf{result\_record\_contract}(Z_i)$, which includes the hash message of recording results, the signature of $M_k$, allocated computational resources, address information, Gas for this transaction, Gas price, etc.
\item Generating blocks of contracts of recording results: Similar to the block generation of contracts of publishing tasks, all members registered in Ethereum can take part in this competition of account-charging right. The winner can obtain Gas as a payment.
\end{enumerate}
In this system, the IoT service acquirer should pay sensors and MEC servers because the original data is from sensors and data process is done by MEC servers. 

\subsection{Communication Model}
The article considers an IRS-assisted uplink communication model with a passive single-antenna eavesdropper (Eve). The IRS is equipped with $N$ programmable phase shifter elements. All channels are assumed to obey Rayleigh fading. The channel from $U_i$ to Eve is defined as $g_{i}\sim\mathcal{CN}(0,1)$,  the channel from IRS to Eve is defined as $\mathbf{z}_{i}^{\rm{H}}\sim\mathcal{CN}_{1,N}(\mathbf{0},\mathbf{I}_N)$, the direct link from $U_i$ to SBS is defined as $l_{i}\sim\mathcal{CN}(0,1)$, the channel from $U_i$ to IRS is defined as $\mathbf{h}_{i}\sim\mathcal{CN}_{N,1}(\mathbf{0},\mathbf{I}_N)$, and the channel from IRS to SBS is defined as $\mathbf{a}_{i}^{\rm{H}}\sim\mathcal{CN}_{1,N}(\mathbf{0},\mathbf{I}_N)$. The instantaneous CSIs of legitimate devices, including $l_i$, $\mathbf{h}_i$, and $\mathbf{a}_i^{\rm{H}}$, can be obtained via channel estimation perfectly, whereas the instantaneous Eve's CSIs $g_{i}$ and $\mathbf{z}^{\rm{H}}_{i}$ are unknown. In this scenario, the TDMA technology is used, so inter-user interference is not existed.

For the security purpose, IRS controls programmable phase shifter elements via a phase shifter matrix, where the phase shifter matrix for the transmission period of $U_i$ is defined as an $N\times N$ matrix $\bm{\Phi}_i$, i.e., 
\begin{flalign}\label{psm}
\bm{\Phi}_i=\text{diag}[\exp(j\theta_{i,1}),...,\exp(j\theta_{i,n}),..., \exp(j\theta_{i,N})],
\end{flalign}
and $\theta_{i,n}\in [0,2\pi)$ is the phase introduced by the $n$th phase shifter element of IRS at the $i$-th period. With the phase shifter matrix $\bm{\Phi}_i$, the received signals at the BS and Eve can be expressed as
\begin{flalign}
& \mathbf{y} = \alpha_i(l_{i}+\mathbf{a}_i^{\rm{H}}\bm{\Phi}_i\mathbf{h}_i)x_i+n_i, \label{mchannel}\\
& \mathbf{y}_{e,i} =\alpha_{e,i}(g_{i}+\mathbf{z}_{i}^{\rm{H}}\bm{\Phi}_i\mathbf{h}_i)x_i+n_{e,i}, \label{Le2}
\end{flalign}
where $\text{diag}(\mathbf{x})$ is the diagonal matrix of $\mathbf{x}$, $\alpha_i$ is the path loss between the SBS and $U_i$, $\alpha_{e,i}$ is the path loss between the SBS and Eve, $x_i$ is the confidential information-bearing signal from $U_i$ with $\mathbb{E}(|x|^2)=P_i$, and $P_i$ is the transmission power of $U_i$. $n_i$ and $n_{e,i}$ are the additive white Gaussian noise (AWGN) obeying $\mathcal{CN}(0,\sigma^2_i)$ and $\mathcal{CN}(0,\sigma_{e,i}^2)$, respectively. 

\subsection{Energy Consumption Model}

The sensor $U_i$ has the computation task $Z_i$ with $D_i$-bits data that should be uploaded to SBS. After receiving the task data, the SBS will select an MEC server, such as $M_k$, to perform the task of $U_i$. In this model, when $D_i$-bits data are calculated in $M_k$, the computing period is $c_kD_i/f_k$, where $c_k$ (CCN/bit) is the CPU cycle number (CCN) per bit processing at $M_k$, and $f_k$ (CCN/s) is the CCN per second of $M_k$. The energy consumption per second of $M_k$ is $\eta_k f_k^3$ (Joule/s), where $\eta_k$ is the computation energy efficiency coefficient of CPU chips in $M_k$. The transmission delay of uploading can be expressed as $D_i/(R_{i}B)$, where $R_i$ (bit/s) is the ergodic secrecy rate of uplink channels, and $B$ is the bandwidth. The energy consumption of $Z_i$ with the assistance of $M_k$ can be formulated as
\begin{flalign}\label{ec} 
Q_{i,k} = \eta_k D_i c_k f_k^2+\frac{D_i P_i}{R_iB}.
\end{flalign}
The computing modules of Ethereum are worldwide distribution, and are independent of the computation offloading system, so the energy consumption of Ethereum is not considered here.

\subsection{Problem Formulation}
Sensors paying higher Gas desire to get better computational resources. Hence, we define the unsatisfactory degree of $U_i$ allocated $M_k$ as follows,
\begin{flalign}\label{unsatisfactory}
O_i=W_{i,k}\bigg[r(v_i)-r\bigg(\frac{f_k}{c_k}\bigg)\bigg],
\end{flalign}
where $W_{i,k}$ is the $i$-th row and $k$-th column of $\mathbf{W}$. $W_{i,k}$ is a binary variable taking 1 when $M_k$ is assigned to $U_i$, and 0 otherwise. $r(v_i)$ is the index of the descending order of $\{v_1,...,v_{N_I}\}$, and $r(f_k/c_k)$ is the index of the descending order of $\{f_k/c_k,...,f_{N_I}/c_{N_I}\}$. Here, $N_K-N_I$ weaker computational resources are abandoned. 

The problem of the computation offloading focuses on the energy consumption minimization with constraints of degrees of satisfaction as follows.
\begin{flalign}
\text{P1: }& \min_{\bm{\Phi}_i,\forall i, \mathbf{W}}\sum_{i=1}^{N_I}\sum_{k=1}^{N_K}W_{i,k}Q_{i,k},\\
& \text{s.t. } O_{i}\leq \epsilon, \forall i, \label{p1c1} \\ 
&\quad \; W_{i,k}= \{0 \text{ or }1\}, \forall i,k, \label{p1c2} \\
&\quad \; \sum_{i=1}^{N_I}W_{i,k}\leq 1, \quad \sum_{k=1}^{N_K}W_{i,k}\leq 1, \label{p1c3} \\
&\quad \; \text{Eq. }(\ref{psm}). \label{p1c4}
\end{flalign}
where Eq. (\ref{p1c1}) requires that the unsatisfactory degree of $U_i$ should be small than a threshold $\epsilon$, and $\epsilon \in [0,1,...,N_I-1]$ that can be adjusted manually. Eq. (\ref{p1c2}) is the binary constraint representing the allocation factor. Eq. (\ref{p1c3}) reveals that each sensor can be allocated with only one MEC server, and each MEC server is only assigned to one sensor. Eq. (\ref{p1c4}) is the passive phase shifter constraint.

It is obvious that $\bm{\Phi}_i,\forall i$ and $\mathbf{W}$ are independent variables in optimization, and P1 is non-convex mixed-integer problem. To tackle the problem, we transform P1 into two sub-problem, i.e., phase shift optimization and computational resource allocation, then solve them step-by-step. 

\section{Computation Offloading Scheme with assistance of IRS and MEC Server}\label{proposed1}

\subsection{Phase Shift Optimization}
From Eq. (\ref{ec}), we can find that the energy consumption $Q_{i,k}$ decreases with the increasing ergodic secrecy rate $R_i$ for all sensors. Firstly, we should find the optimal $\bm{\Phi}_i$ to maximize $R_i$, where $R_i$ is given as follows \cite{Liu2020},
\begin{flalign}\label{esr}
R_i& = [\mathbb{E}(C_{m,i})-\mathbb{E}(C_{w,i})]^+ \\ 
&\leq \mathbb{E}[(C_{m,i}-C_{w,i})^{+})], \label{realesr}
\end{flalign}
where
\begin{flalign}\label{esr}
&C_{m,i}=B\log_2\bigg(1+\frac{\alpha_i^2P_i}{\sigma_i^2}|l_{i}+\mathbf{a}_i^{\rm{H}}\bm{\Phi}_i\mathbf{h}_i|^2\bigg), \\ 
&C_{w,i}=B \log_2\bigg(1+\frac{\alpha_{e,i}^2P_i}{\sigma_{e,i}^2}|g_{i}+\mathbf{z}_{i}^{\rm{H}}\bm{\Phi}_i\mathbf{h}_i|^2\bigg),
\end{flalign}
and $B$ is the bandwidth. $\{=\}$ in Eq. (\ref{realesr}) holds if and only if the  instantaneous
secrecy rate $\{C_{m,i}-C_{w,i}\}$ is nonnegative in all channel state. Due to Eve's CSI $g_{i}$ and $\mathbf{z}$ are unknown, it is hard to determine whether an instantaneous secrecy rate is nonnegative or not, so we use the lower bound of real ergodic secrecy rate as the performance metric for the optimization process. The objective is to find the optimal $\bm{\Phi}_i$ to achieve $R_i^*=\max_{\bm{\Phi}}R_i$ and get the expression of $R_i^*$ for the following computational resource allocation. However, it is hard to maximize $R_i$ as $g_{i}$ and $\mathbf{z}$ are unknown, so the objective is transformed to maximize the channel capacity between $U_i$ and the SBS as follows, 
\begin{flalign}
\text{P2: }& \max_{\bm{\Phi}}\mathbb{E}(C_{m,i}), \quad \text{s.t. } \text{Eq. }(\ref{psm}). \label{p2c1}
\end{flalign}
From the investigation in \cite{Wu2019}, the optimal $\bm{\Phi}_i$ of P2, i.e., $\bm{\Phi}_i^*$, can be found as follows, 
\begin{flalign}\label{thetaopt}
\theta_{i,n}^*=\theta_i-\text{arg}(a^{\rm{H}}_{i,n})-\text{arg}(h_{i,n}), n=1,...,N,
\end{flalign}
where $\theta_{i,n}^*$ is the $n$-th diagonal element in $\bm{\Phi}_i^*$, $\theta_i=\text{arg}(l_i)$, $a^{\rm{H}}_{i,n}$ is the $n$-th element of $\mathbf{a}^{\rm{H}}_i$, and $h_{i,n}$ is the $n$-th element of $\mathbf{h}_i$. Here, $\text{arg}(x)$ is the angle of a complex variable $x$. 

\begin{theorem}[Expression of optimal ergodic secrecy rate]
The expression of optimal ergodic secrecy rate of $U_i$ with $\bm{\Phi}^*_i$ can be expressed as
\begin{flalign}\label{eesr}
&R_i^*=[\mathbb{E}(C_{m,i}|\bm{\Phi}^*)-\mathbb{E}(C_{w,i}|\bm{\Phi}^*)]^+,
\end{flalign}
where 
\begin{flalign}\label{cm}
&\mathbb{E}(C_{m,i}|\bm{\Phi}^*)\\
&=\frac{2^{\mu_1-1/2}}{\ln(2)\Gamma(\mu_1)\sqrt{2\pi}}G_{3,5}^{5,1}\bigg( \frac{\sigma^2}{4\nu_1^2P_i\alpha_i^2}\bigg| \begin{matrix}0, \frac{1}{2}, 1 \\
0,0,\frac{1}{2},\frac{\mu_1}{2}, \frac{\mu_1+1}{2}
\end{matrix} \bigg),
\end{flalign}
\begin{flalign}
\mu_1=\frac{(\sqrt{\pi}+2\eta\kappa N)^2}{4+4\eta\kappa^2 N-\pi}, \quad \nu_1 = \frac{4+4\eta\kappa^2-\pi}{2(\sqrt{\pi}+2\eta\kappa N)},
\end{flalign}
$\eta=\pi^2/(16-\pi^2)$, and $\kappa=(4-\pi^2/4)\pi$. 
\begin{flalign}\label{cw}
\mathbb{E}(C_{w,i}|\bm{\Phi}^*)=\frac{1}{\ln(2)\Gamma(\mu_2)}G_{2,3}^{3,1}\bigg( \frac{\sigma_{e,i}^2}{\nu_2 P_i \alpha_{e,i}^2 }\bigg| \begin{matrix}0, 1 \\
0,0, \mu
\end{matrix} \bigg),
\end{flalign}
where 
\begin{flalign}
\mu_2=\frac{(1+N)^2}{(1+N)^2+2N}, \quad \nu_2 = 1+N+\frac{2N}{1+N}.
\end{flalign}

\end{theorem}

\vspace{0.2cm}

\begin{proof}
The similar proof of Eqs. (\ref{cm}) and (\ref{cw}) can be found in \cite[Eq. (21)]{VanChien2021} and \cite[Eq. (20)]{VanChien2021}, respectively. Substituting Eqs. (\ref{cm}) and (\ref{cw}) into Eq. (\ref{esr}), we can obtain the expression of optimal ergodic secrecy rate.
\end{proof}

\subsection{Gas-Oriented Computational Resource Allocation}
The optimal computational resource allocation is found by grouping and matching algorithms.

\subsubsection{Grouping process} In order to meet the satisfactory degree of each sensor, i.e., constraint (\ref{p1c1}), the sensors are sort as $\{ U_{1'},U_{2'},...,U_{N_I'}\}$ where their Gas obey $V_{1'}\geq V_{2'} \geq ... \geq V_{N_I'}$, and are grouped into $T=\lceil \frac{N_I}{\epsilon+1} \rceil$ groups. Each group includes $\epsilon+1$ members as follows, 
\begin{flalign}
&\overbrace{\{ U_{1'},U_{2'},...,U_{(\epsilon+1)'}\}}^{\text{first sensor group}}, \quad \overbrace{\{ U_{(\epsilon+2)'},U_{(\epsilon+3)'},...,U_{(2\epsilon+2)'}\}}^{\text{second sensor group}},\notag \\
&..., \overbrace{\{ U_{[(T-1)\epsilon+T]'},U_{[(T-1)\epsilon+T+1]'},...,U_{N_I'}\}}^{\text{last sensor group}}.
\end{flalign}
Specially, the last sensor group has $\{N_I-(T-1)(\epsilon+1)\}$ members. Similarly, the MEC servers are sort as $\{ M_{1''},M_{2''},...,M_{N_I''}\}$ where their computational power obey $f_{1''}/c_{1''}\geq f_{2''}/c_{2''} \geq ... \geq f_{N_I''}/c_{N_I''}$ by dropping $N_K-N_I$ weaker MEC servers, and are grouped into $T$ groups. Each group includes $\epsilon+1$ members as follows, 
\begin{flalign}
&\overbrace{\{ M_{1''},U_{2''},...,M_{(\epsilon+1)''}\}}^{\text{first MEC group}}, \quad \overbrace{\{ M_{(\epsilon+2)''},M_{(\epsilon+3)''},...,M_{(2\epsilon+2)''}\}}^{\text{second MEC group}},\notag \\
&..., \overbrace{\{ M_{[(T-1)\epsilon+T]''},M_{[(T-1)\epsilon+T+1]''},...,M_{N_I''}\}}^{\text{last MEC group}}.
\end{flalign}
The last MEC group has $\{N_I-(T-1)(\epsilon+1)\}$ members. The MEC server in the $t$-th MEC group is only allocated to the $t$-th sensor group, and the sensor in the $t$-th sensor group can only uses computational resources in the $t$-th MEC group. From Eq. (\ref{unsatisfactory}), we can find that $r(v_i)-r(f_k/c_k)\leq \epsilon$ for any sensor and MEC server pair, so the constraint (\ref{p1c1}) is satisfied.

\begin{figure*}[htb]
\centering
\begin{minipage}{.32\textwidth}
\centering
\includegraphics[width=0.94\linewidth]{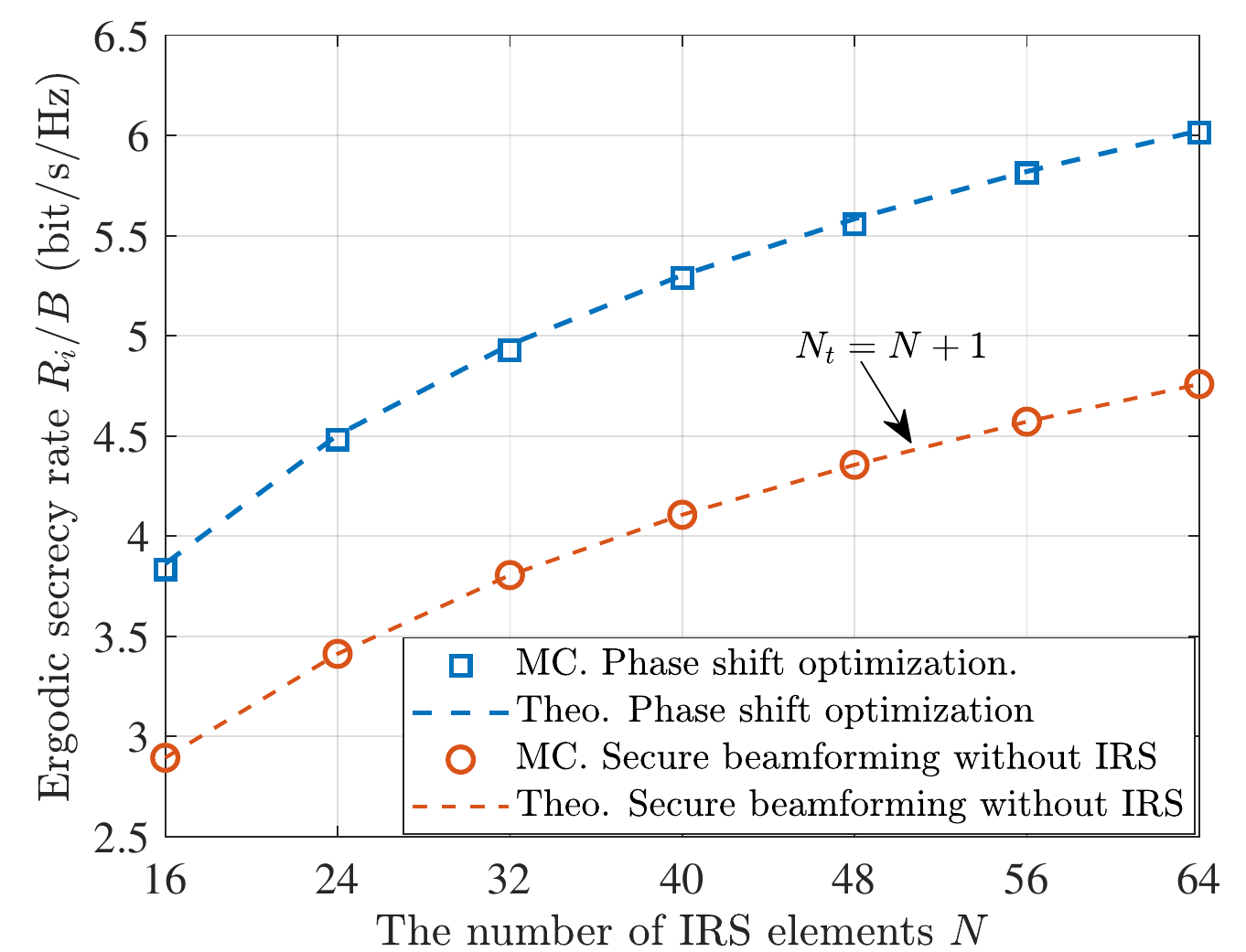}
\caption{Optimal ergodic secrecy rate of $U_i$ in terms of the number of IRS elements.}
\label{sim1}
\end{minipage}
\hfil
\begin{minipage}{0.32\textwidth}
\centering
\includegraphics[width=0.94\linewidth]{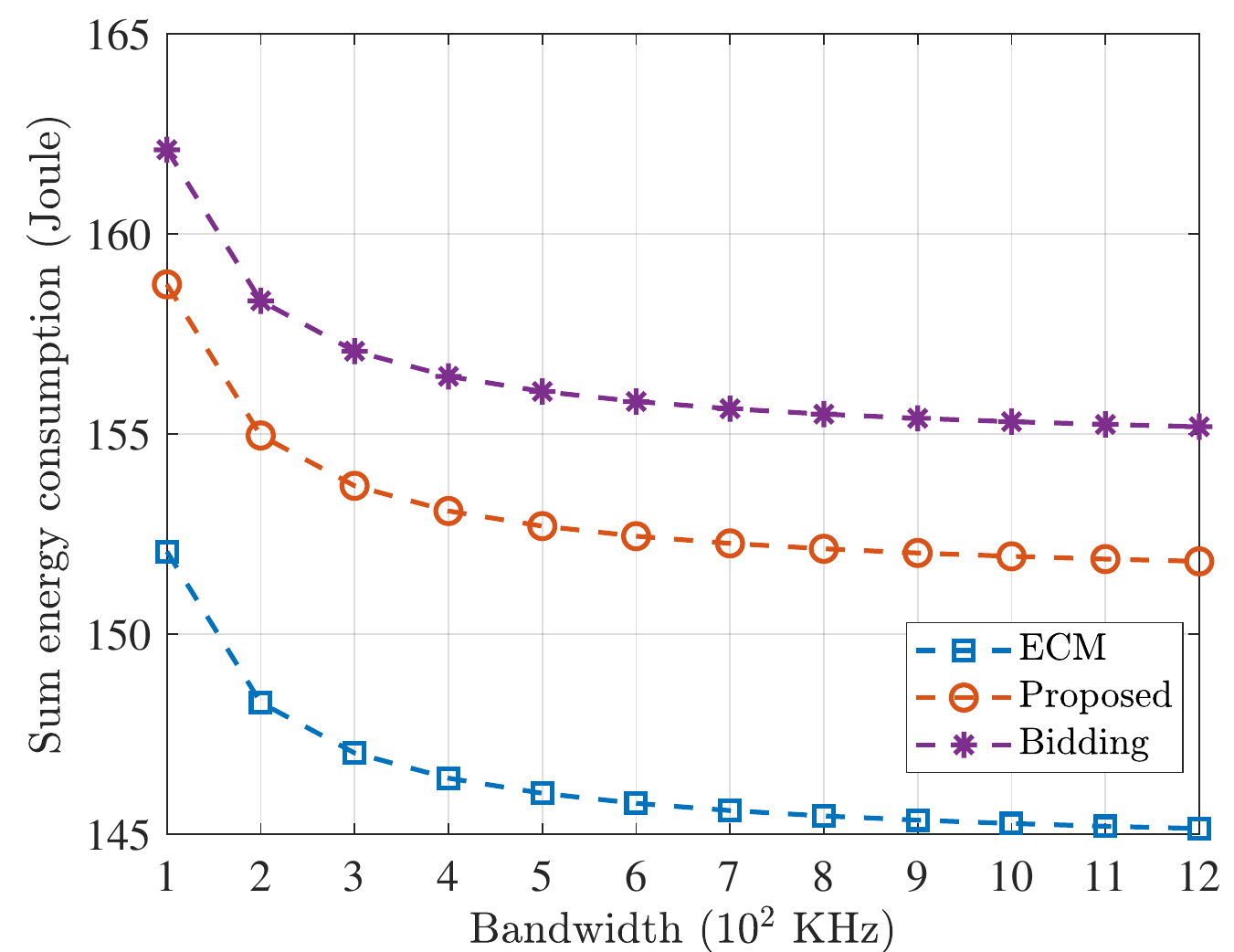}
\caption{Sum energy consumption of 40 sensors in terms of the length of bandwidth.}
\label{sim2}
\end{minipage}
\hfil
\begin{minipage}{0.32\textwidth}
\centering
\includegraphics[width=0.94\linewidth]{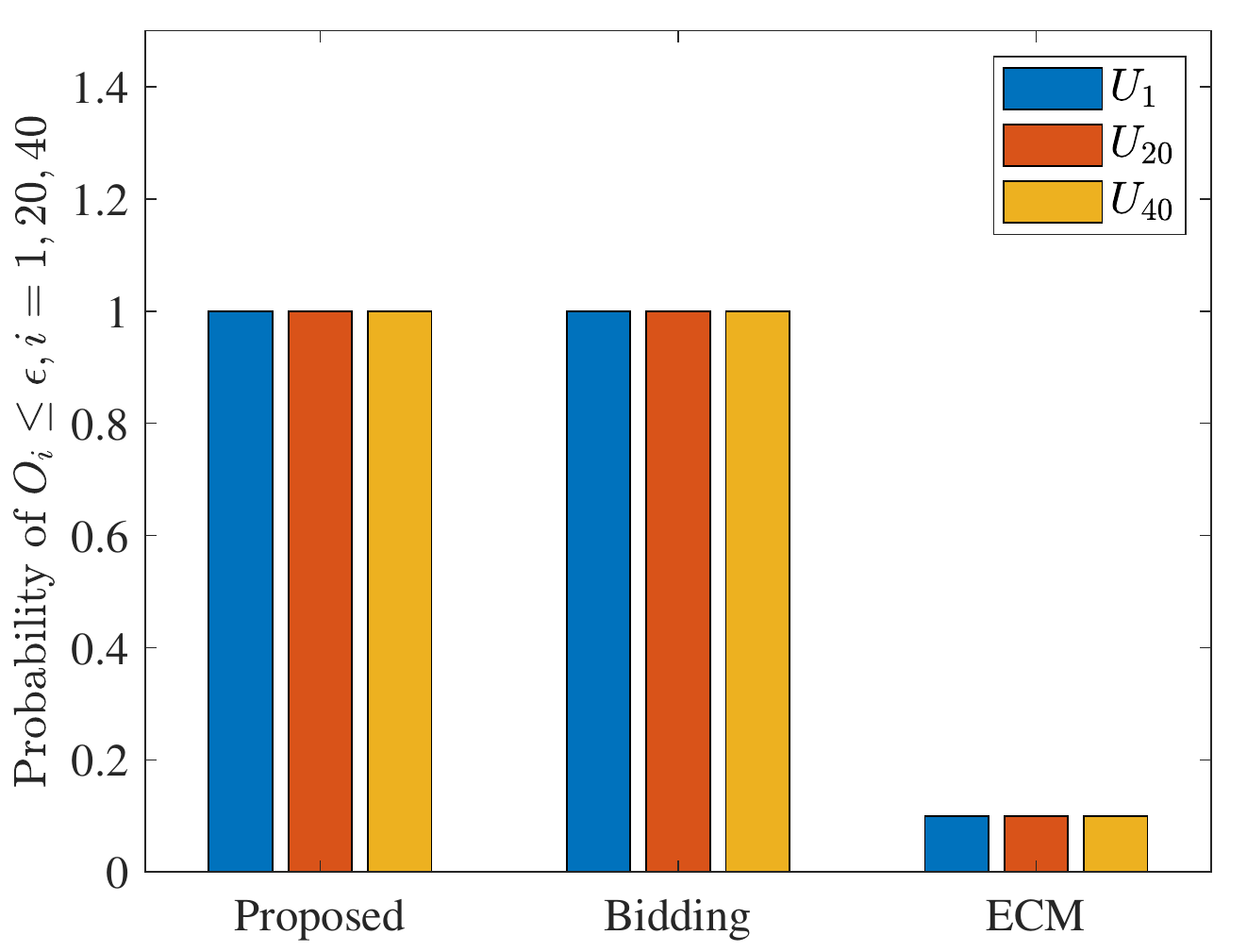}
\caption{The probability of satisfaction of $U_1$, $U_{20}$, and $U_{40}$.}
\label{sim3}
\end{minipage}
\end{figure*}

\subsubsection{Matching process} Without loss of generality, we define $\mathcal{S}_t=\{[(t-1)\epsilon+t]',...,t(\epsilon+1)'\}$ and $\mathcal{M}_t=\{[(t-1)\epsilon+t]'',...,t(\epsilon+1)''\}$ as the indices of the members in the $t$-th sensor group and the $t$-th MEC group. Then, we use Theorem 1 to calculate optimal ergodic secrecy rates of all sensors, i.e., $R_i^*,\forall i\in \mathcal{S}_t$. With the parameters $\{\eta_k,c_k,f_k,D_i,P_i,R_i^*\},\forall i \in \mathcal{S}_t, \forall k\in \mathcal{M}_t$, we can calculate $Q_{i,k}$ via Eq. (\ref{ec}) for each given sensor and MEC server pair, and generate a matrix $\mathbf{Q}$ to record $Q_{i,k}, \forall i\in\mathcal{S}_t, \forall k\in \mathcal{M}_t$. The allocation problem in the $t$-th sensor group and the $t$-th MEC group can be equivalently transformed to a 2-dimensional matching problem as 
\begin{flalign}
\text{P3: }& \min_{\mathbf{W}_t}\sum_{i=(t-1)\epsilon+t}^{\epsilon+1}\sum_{k=(t-1)\epsilon+t}^{\epsilon+1} W_{t,i,k}Q_{i,k}, \\
& \text{s.t. } W_{t,i,k}= \{0 \text{ or }1\}, \forall i,k, \label{p3c1} \\
&\quad \; \sum_{i=(t-1)\epsilon+t}^{\epsilon+1}W_{t,i,k}\leq 1, \; \sum_{k=(t-1)\epsilon+t}^{\epsilon+1}W_{t,i,k}\leq 1, \label{p3c2}
\end{flalign}
where $W_{t,i,k}$ is the $i$-th row and $k$-th column of $\mathbf{W}_t$. $W_{t,i,k}$ is a binary variable taking 1 when $M_k$ in $t$-th MEC group is assigned to $U_i$ in the $t$-th sensor group, and 0 otherwise. Eq. (\ref{p3c2}) reveals that each sensor can be allocated with only one MEC server, and each MEC server is only assigned to one sensor. P3 is convex and the optimal $\mathbf{W}_t$ can be solved by the Kuhn-Munkres (KM) algorithm \cite{inproceedings}. Even if the number of dimensional in the last group is different from the formers, computational resource allocation in all $T$ sensor and MEC groups, including the last group, can be optimized via the KM algorithm.

\subsection{Computational Complexity Analysis}

In the phase optimization process, Since Eq. (\ref{thetaopt}) requires $2N$ iterations, the computational complexity to get $\bm{\Phi}_i^*, \forall i$ is $O(2NN_{I})$. The computational resource allocation requires two bubble sort algorithms to get $\{ U_{1'},U_{2'},...,U_{N_I'}\}$ and $\{ M_{1''},M_{2''},...,M_{N_I''}\}$, each of which needs $N_I^2$ iterations. The KM algorithm needs $(\epsilon+1)^4$ iterations in each group pair \cite{inproceedings}.  In total, the computational complexity of the offloading process is $O[2NN_{I}+2N_I^2+T(\epsilon+1)^4]$.

\section{Simulations}\label{simulations}
The global simulation parameters are described as follows. The path loss parameter $\alpha_i$ is calculated by $\alpha_i=\frac{c}{2\pi f_c d_i}$, where $c$ is the speed of light, $f_c$ is work spectrum that is set to 2.4 GHz, $d_i$ is the distance between the SBS and $U_i$ is set to be uniform distribution over $[$30 m, 50 m$]$. The AWGN floor parameters $\sigma_i^2$ and $\sigma_{e,i}^2$ are -53 dBm \cite{Dinc2016}. The transmission power $P_i, \forall i$ is 10 dBm. The computation energy efficiency coefficient $\eta_k, \forall k$ is set to be $10^{-27}$. The CCN per bit processing $c_k, \forall k$ is set to be 10 CCN/bit. Note that all simulation results are average values from $10^5$ independent runs.

Here, we examine Theorem 1 in terms of the number of IRS elements in Fig. \ref{sim1}. These figures show the good agreements between theoretical results (Theo.) of Eq. (\ref{eesr}) and Monte Carlo (MC.) simulation results of Eq. (\ref{esr}) from $10^5$ independent runs. From this figure, we can find that the optimal ergodic secrecy rate increases with the increasing number of IRS elements. Also, a comparison is taken between IRS-assisted PLS and the secure beamforming scheme that uses $N_t=N+1$ antennas \cite{Liu2020}. From the comparison simulations, we find that the IRS-assisted PLS outperforms the secure beamforming scheme because channel gain by IRS is larger than that of the beamforming scheme \cite{VanChien2021}. 

Fig. \ref{sim2} shows the bandwidth effect on the sum energy consumption of 40 sensors, where $\epsilon=9$. $f_k$, $D_i$, and $V_i$ are uniform distributions over $[$40 GHz, 80 GHz$]$, $[$610 KB 1.8 MB$]$, and $[1.5\times 10^6, 2\times 10^6]$ . It is demonstrated that the sum energy consumption decreases with the increasing bandwidth, because data transmission time is reduced with more bandwidth. Also, the proposed scheme outperforms the bidding scheme (the highest bidder obtains), but has worse performance than energy consumption minimization (ECM) schemes \cite{Wang2016,Feng2020}. From the red and blue lines, we can find the proposed scheme just raises $4\%$ enough consumption compared to ECM. Fig. \ref{sim3} with the simulation parameters of Fig. \ref{sim2} is used to check whether sensors are satisfied. We can find the probabilities of the satisfaction of the proposed scheme and bidding scheme is equal to one, meaning that sensors are satisfied. However, the provided Gas has no relationship with computational resource allocation in the ECM schemes, so the probability of $O_i\leq \epsilon$ is equal to $1/T$, where $T=N_I/(\epsilon+1)=10$ in this simulation.

\section{Conclusions}\label{conclusions}
In this article, we propose an endogenous security architecture for computation offloading in IoT, which not only provides the ability to trace unknown malicious behaviors of nodes by blockchain technology, but also uses PLS to protect the uplink channel from sensors to MEC servers against passive eavesdropping attacks. The feature of the proposed scheme is that the group with higher Gas is prioritized with better computational resources. The simulations demonstrate that the proposed scheme guarantees that the node paying higher Gas has more opportunities to get a stronger computational resource, and just raises $4\%$ energy consumption in comparison with ECM schemes. In future works, we will integrate multiple antenna technologies into this system as it is a promising method to improve security performance.

\bibliographystyle{IEEEtran}
\bibliography{references}
\end{document}